\documentclass{article}%
\usepackage{amsmath}
\usepackage{graphicx}
\usepackage{amsfonts}
\usepackage{amssymb}%
\providecommand{\U}[1]{\protect\rule{.1in}{.1in}}

\ifx\pdfoutput\relax\let\pdfoutput=\undefined\fi
\newcount\msipdfoutput
\ifx\pdfoutput\undefined\else
\ifcase\pdfoutput\else
\ifx\paperwidth\undefined\else
\ifdim\paperheight=0pt\relax\else\pdfpageheight\paperheight\fi
\ifdim\paperwidth=0pt\relax\else\pdfpagewidth\paperwidth\fi
\fi\fi\fi
\begin{document}

\begin{flushright}
MZ-TH/10-37

September 2013
\end{flushright}

\begin{center}
\medskip{\LARGE New Sum Rule Determination of the Mass and Strangeness
Content}

{\LARGE of the Nucleon\medskip\ }
\end{center}

\vspace{2cm}

\begin{center}
{\large N.F. Nasrallah}$^{(a)}${\large \ and K. Schilcher}$^{(b),(c)}$

$^{a}$Lebanese University, Faculty of Science

Tripoli, Lebanon\newline$^{(b)}$ Institut f\"{u}r Physik, Johannes
Gutenberg-Universit\"{a}t\newline Staudingerweg 7, D-55099 Mainz, Germany

$^{(c)}$Centre for Theoretical Physics and Astrophysics\newline University of
Cape Town, Rondebosch 7700, South Africa
\end{center}

\vspace{0.2cm}{\footnotesize \textit{E-mail:} nsrallh@ul.edu.lb,
karl.schilcher@uni-mainz.de }

\begin{center}
\textbf{Abstract\medskip}
\end{center}

\begin{quote}
A new QCD calculation of the mass of the nucleon is presented. It makes use of
a polynomial kernel in the dispersion integrals tailored to practically
eliminate the contribution of the unknown $1/2^{+}$ and $1/2^{-}$ continuum.
This approach avoids the arbitrariness and instability attached to the Borel
kernel used in previous sum rules calculations. Our method yields stable
results for the nucleon mass and coupling. For standard values of the
condensates, the prediction of the nucleon mass in the chiral limit is
$m_{N}=(830\pm50)MeV$. With the pion-nucleon sigma-term given by chiral
perturbation theory and the strange sigma-term estimated by the Zweig rule we
get \ $m_{N}=(990\pm50)MeV.$

\bigskip
\end{quote}

{\newpage}

\section{Introduction{}}

The nucleon sigma terms are defined by
\[
\sigma_{\pi N}=\hat{m}\left\langle N(p)|\overline{u}u(0)+\overline
{d}d(0)|N(p)\right\rangle \
\]%
\[
\sigma_{s}=m_{s}\left\langle N(p)|\overline{s}s(0)|N(p)\right\rangle \
\]
where $\hat{m}=\frac{1}{2}(m_{u}+m_{d})$ and $m_{u},m_{d},m_{s}$ refer to
quark masses. The sigma term was introduced in chiral perturbation theory
(ChPT) to measure the explicit breaking of chiral symmetry due to non-zero
masses of light quarks .It represents the contribution from the finite quark
masses to the mass of the nucleon and it contains important information on the
strangeness content of the nucleon and the quark mass ratios. The pion-nucleon
sigma term is related to the value of the pion-nucleon invariant amplitude at
the unphysical Cheng-Dashen point where $s-u=0,t=2m^{2}$ (here, s, t, u are
the Mandelstam variables). Recently in a paper on the elastic scattering of
supersymmetric cold dark matter particles on nucleons it has been shown that
the cross sections depend strongly on the value of the pion-nucleon sigma term
$\sigma_{\pi N}$\cite{Ellis}. Related to the sigma terms is the so-called
strangeness content of the nucleon be defined as%
\[
y\equiv2\frac{\left\langle N(p)|\overline{s}s|N(p)\right\rangle }{\left\langle
N(p)|\overline{u}u+\overline{d}d|N(p)\right\rangle }\ .
\]
The quantity $y$ relates the two sigma terms,%
\begin{equation}
\sigma_{s}=\frac{y}{2}\frac{m_{s}}{\hat{m}}\sigma_{\pi N}\text{ .} \label{2}%
\end{equation}
The sigma term can be calculated from Lattice Gauge Theory (LGT) using the
Feynman-Hellmann theorem applied to the nucleon mass. The LGT results are
rather frustrating. Predictions vary between $20MeV$ and $90MeV$, with
incompatible errors (for a recent compilation of LGT results see \cite{Bali}).
Experimental results come from dispersion relation analysis of pion-nucleon
scattering data. They are equally confusing. To quote two recent
publications.:A George Washington University phase-shift analysis from 2005
leads to a sigma term of $81\pm6MeV$ \cite{Hite}. A more recent analysis of
the same TRIUMF data\cite{Stahov} gives $\sigma_{N}=43\pm12MeV$.

In view of the importance of the sigma term the situation concerning both
theory and experiment is totally frustrating. We therefore hope to put some
light into the issue by communicating our results of a QCD sum rule analysis
of the nucleon mass which allows conclusions on the sigma terms and the
strangeness content of the nucleon. The QCD sum rule method introduced by
Shifman et al. \cite{1} has extended the applicability of QCD far beyond
simple perturbation theory. The method was adapted to the case of nucleons by
Ioffe \cite{2} and independently by Chung, Dosch, Kremer and Schall \cite{3}.
These authors showed how to approach one of the fundamental problems of QCD,
the calculation of baryon masses from the Lagrangian and the vacuum
condensates. The critical problem of previous calculations was the unknown
couplings of the higher nucleon resonances to the nucleonic current on the
hadronic side of the sum rule. Using a new sum rule approach, we obtain for
the first time stable and unambiguous results for the nucleon mass in the
chiral limit.

Nucleon mass sum rules start with the correlation function
\begin{equation}
\Pi(q)=i\int d^{4}xe^{iqx}\left\langle 0|\eta(x)\eta(0)|0\right\rangle
\label{6}%
\end{equation}
where $\eta$ is a nucleon interpolating field constructed from local QCD
operators with the quantum numbers of the nucleon. We will choose \cite{2}
\[
\eta=e^{abc}(u^{a}C\gamma^{\lambda}u^{b})\gamma_{5}\gamma^{\lambda}d^{c}\ .
\]
which couple maximally to the nucleon. The correlator can be decomposed in
terms of invariants,
\[
\Pi(q)=q_{\mu}\gamma^{\mu}\Pi_{1}(q^{2})+\Pi_{2}(q^{2})
\]
with $\gamma_{\mu}$ standing for the Dirac matrices. $\Pi(t=q^{2})$ is an
analytic function in the complex $t$-plane with a pole at $t=m_{N}^{2}$ and a
cut along the positive real axis starting at $t=(m_{N}+m_{\pi})^{2}$. The sum
rule methods can be traced back to the Cauchy formula
\begin{equation}
\frac{1}{2\pi i}\oint\Pi(t)P(t)\,dt=-\int_{0}^{R}\frac{1}{\pi}\text{Im}%
\Pi(t)P(t)\,dt \label{7}%
\end{equation}
where the kernel $P(t)$ is an arbitrary analytic function. The integral on the
left hand side (l.h.s.) is over a circle of radius $R$. If $R$ is taken large
enough, we can replace $\Pi(t)$\ on the l.h.s. by it's QCD and operator
product expansion (OPE) counterpart $\Pi_{QCD}(t).$ The right hand side
(r.h.s.) involves, apart from the nucleon pole, an integral over the cut,
consisting of a background plus a set of nucleonic resonances. Duality means,
that the OPE result on the l.h.s. of eq. \ref{7}, is equated to the hadronic
contribution on the right hand side. Traditionally the integrand on the r.h.s.
is approximated by the \textquotedblleft pole\ plus
continuum\textquotedblright\ model,
\begin{equation}
\frac{1}{\pi}\operatorname{Im}\Pi(q^{2})=\lambda^{2}\delta(q^{2}-m_{N}%
^{2})+\frac{1}{\pi}\theta(q^{2}-W^{2})\operatorname{Im}\Pi^{OPE}(q^{2})
\label{8}%
\end{equation}
Here $m_{N}$ is the position of the lowest lying pole with residue
$\lambda_{N}$, the coupling of the current to the nucleon state
\[
\left\langle 0|\eta|n\right\rangle =\lambda\Psi\ ,
\]
and an effective continuum threshold $W^{2}$ which is determined in the
calculation and on which the results depend sensitively.

Most sum rule studies of baryonic currents invoke a Borel transform of the
correlator, i.e. they use a kernel
\begin{equation}
P(t)=e^{-t/M^{2}} \label{9}%
\end{equation}
which introduces another, more or less arbitrary, parameter providing
exponential damping of the continuum(when it is small) and suppressing high
dimensional vacuum condensates(when it is large).The integral over the circle
of the non-perturbative part amounts to an infinite sum of terms of the form
$\frac{C_{n}}{n!M^{2n}}$ which become important when $M$ is small. Stability
has to be established under variations of the latter parameters.

The arbitrariness in the choice of the parameters $W^{2}$ and $M^{2}$ makes
that the method results in an estimate rather than a calculation of the
nucleon mass.

To overcome these intrinsic ambiguities we have introduced some time ago a sum
rule method\ \cite{ACD}, originally called ACD, which exploits the analyticity
properties of the correlator to significantly reduce, in some cases
practically eliminate, the contribution of the continuum. The breakthrough in
the treatment of the continuum has been the introduction of an integration
kernel in the FESR tuned to suppress substantially the resonance energy region
above the ground state. This approach, specially adapted to eliminate
pronounced resonances, has been recently used to extract very precise values
of the light quark masses \cite{CAD09} and condensates \cite{Bordes} as well
as an evaluation of the neutron-proton mass difference \cite{Nasrallah2013}.
Our approach is based on the fact, that the contribution of the continuum in
the integral on the r.h.s. of eq.(\ref{7}) arises mostly from the interval
\begin{equation}
I=2.0GeV^{2}\leq t\leq3.0GeV^{2} \label{11}%
\end{equation}
where the four nucleon resonances $N^{+}(1440),\ N^{-}(1535),\ N^{-}(1650)$
and $N^{+}(1710)$ lie. This prompts us to choose
\begin{equation}
P(t)=1-a_{1}t-a_{2}t^{2} \label{12}%
\end{equation}
The parameters $a_{1}$and $a_{2}$ are chosen so as to minimize the integral
$\int_{2GeV^{2}}^{3GeV^{2}}\left\vert P(t)^{2}\right\vert dt$. Numerically
$a_{1}=.807GeV^{-2}$ and $a_{2}=-.160GeV^{-4}$. With this choice the relative
damping over the interval $I$ , $\left\vert P(t)/P(m_{N}^{2})\right\vert $
does not exceed 6\% (see Fig.1). The corresponding quantity in the case of
exponential kernel (for $M^{2}=1.1GeV^{2}$ e.g. ) is $36\%$ which shows that
our choice, eq.(\ref{11}) provides considerably better damping to the
contribution of the continuum and justifies its neglect. Another advantage of
our choice is that it will involve the contribution of only one higher order
unknown condensate in the calculation whereas the exponential kernel involves
an infinite number of these. Similar damping of the continuum is obtained for
the kernel \thinspace$tP(t)$. A residual model dependency is still unavoidable
as inelasticity, non-resonant background and resonance interference are
impossible to guess realistically. Also here our approach helps, as a constant
background is eliminated by the integration kernel. Having thus minimized the
contribution we will neglect it.%
{\begin{center}
\includegraphics{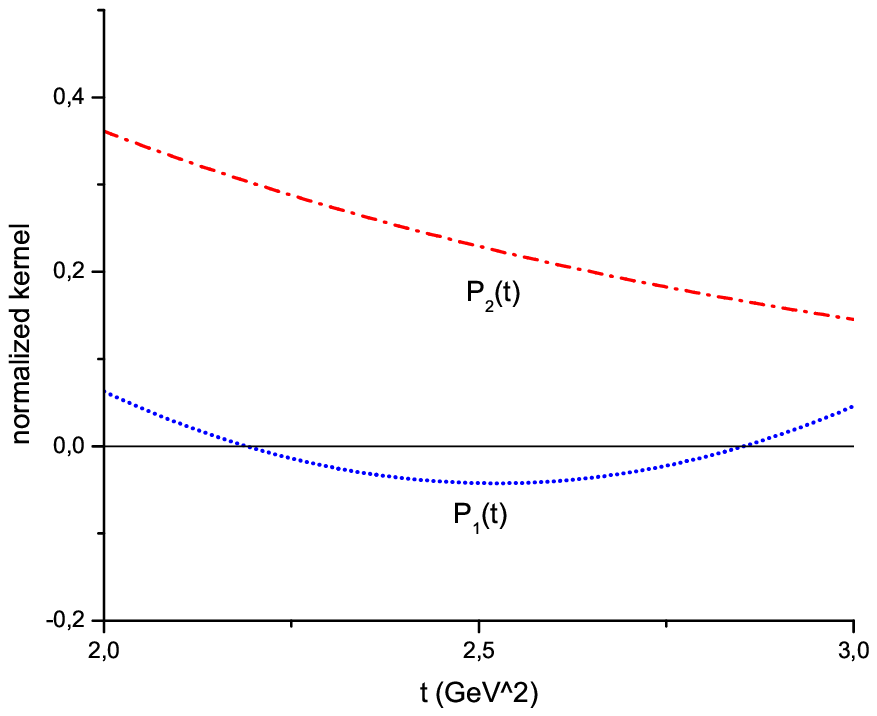}
\\
Fig. 1: $P_1(t)$ is our polynomial of Eq.(\ref{12}) and $P_2(t)$ is the Borel
kernel of Eq.(\ref{9}). Both kernels are normalized to $1$ at the nucleon
mass.
\end{center}}

The theoretical side of the sum rule, in contrast, is in better shape. The
correlator (\ref{6}) is known including radiative corrections and OPE terms up
to dimension d = 9 \cite{2},\cite{Jamin},\cite{Sadovnikova}. There exists the
usual uncertainty about the precise values of the condensates and the validity
of the factorization assumption used for the higher dimensional condensates.
We note that the kernel eq.(\ref{12}) will introduce only low dimension
condensates into the calculation which are known (or at least estimated).

Apart from the references cited above there are a few more attempts to
evaluate the sum rule (\ref{7}) involving a high sophistication on the
theoretical side which is not always commensurate with the primitive model
ansatz on the phenomenological side. We therefore think it necessary to
present for once an (almost) model independent investigation of the nucleonic
sum rule.

\section{The calculation}

The invariant amplitudes have poles at $t=m_{N}^{2}$%

\begin{align*}
\Pi_{1}(t)  &  =\frac{-\lambda^{2}}{(t-m_{N}^{2})}+...\\
\Pi_{2}(t)  &  =\frac{-m_{N}.\lambda^{2}}{(t-m_{N}^{2})}+...
\end{align*}

It follows then from Cauchy's theorem that

\bigskip%

\begin{align*}
|\lambda|^{2}P(m_{N}^{2})  &  =\frac{-1}{2\pi i}\oint\limits_{|t|=R}dt\Pi
_{1}^{QCD}P(t)+\frac{1}{\pi}\int_{th}^{R}dtP(t)\operatorname{Im}\Pi_{1}(t)\\
m_{N}|\lambda|^{2}P(m_{N}^{2})  &  =\frac{-1}{2\pi i}\oint\limits_{|t|=R}%
dt\Pi_{2}^{QCD}P(t)+\frac{1}{\pi}\int_{th}^{R}dtP(t)\operatorname{Im}\Pi
_{2}(t)
\end{align*}
with%

\begin{align}
(2\pi)^{4}\Pi_{1}^{QCD}(t)  &  =A_{0}t^{2}\ln\frac{-t}{\mu^{2}}+A_{01}%
t^{2}\left(  \ln\frac{-t}{\mu^{2}}\right)  ^{2}+A_{4}\ln\frac{-t}{\mu^{2}%
}\nonumber\\
&  +A_{6}\frac{1}{t}++A_{61}\frac{1}{t}\ln\frac{-t}{\mu^{2}}+A_{8}\frac
{1}{t^{2}}+.... \label{3}%
\end{align}
and
\begin{equation}
(2\pi)^{4}\Pi_{2}^{QCD}(t)=B_{3}t\ln\frac{-t}{\mu^{2}}+B_{7}\frac{1}{t}%
+B_{9}\frac{1}{t^{2}}+... \label{4}%
\end{equation}

The coefficients $A_{i}$ and $B_{i}$ are defined as in \cite{Sadovnikova} but
for a factor $(2\pi)^{4}$and powers of $t$ taken explicitly%

\begin{align}
A_{0}  &  =-\frac{1}{4}(1+\frac{71}{12}a),A_{01}=\frac{a}{8}\nonumber\\
A_{4}  &  =-\frac{\pi^{2}}{2}\left\langle aGG\right\rangle \nonumber\\
A_{6}  &  =-\frac{2}{3}(2\pi)^{4}\left\langle \bar{q}q\bar{q}q\right\rangle
(1-\frac{5}{6}a),A_{61}=\frac{2}{9}(2\pi)^{4}\langle qqqq\rangle a\nonumber\\
A_{8}  &  =\frac{-1}{6}(2\pi)^{4}\mu_{0}^{2}\left\langle \bar{q}q\bar
{q}q\right\rangle \nonumber\\
B_{3}  &  =4\pi^{2}\left\langle \bar{q}q\right\rangle \left(  1+\frac{3}%
{2}a\right) \nonumber\\
B_{7}  &  =-\frac{4\pi^{4}}{3}\left\langle \bar{q}q\right\rangle \left\langle
aGG\right\rangle \nonumber\\
B_{9}  &  =-(2\pi)^{6}\frac{136}{81}a\left\langle (\bar{q}q)^{3}\right\rangle
\ \ \ \ \text{\ }\nonumber
\end{align}
where $a=\frac{\alpha_{s}(\mu^{2})}{\pi}$. The terms $B_{7}$ and $B_{9}$ are
given in the factorization approximation. In $A_{8}$ we have taken
\[
\left\langle 0\right\vert \bar{q}q\bar{q}aG_{\mu\nu}^{c}\frac{\lambda_{c}}%
{2}\sigma^{\mu\nu}q\left\vert 0\right\rangle =\mu_{0}^{2}\left\langle \bar
{q}q\bar{q}q\right\rangle
\]
with the parameter $\mu_{0}^{2}=0.8GeV^{2}$ as advocated in \cite{Belyaev}..
The question at which scale this relation holds is resolved by allowing for
generous errors. To avoid the double counting, we keep the logarithmic
$\ln(-t/\mu^{2})$ contribution in the polarization operator but neglect its
anomalous dimension. In any case anomalous dimension effects are very small
\cite{Pivovarov}.

The contribution of the radiative corrections $A_{01}$ and $A_{06}$ turn out
to be smaller than the estimated errors so we discard them for simplicity.

For the finite energy sum rule eq.(\ref{7}), we need the known integrals of
the form
\[
I_{ik}=\frac{1}{2\pi i}\oint dtt^{i}\left(  \ln(-t)\right)  ^{k}%
\]
These are given in convenient form in \cite{Bordes}.

With our choice of $P(t)$, we then get
\begin{equation}
(2\pi)^{4}|\lambda|^{2}P(m_{N}^{2})=-A_{0}I_{2}(R)-A_{4}I_{0}(R)-A_{6}%
+a_{1}A_{8}+\Delta_{1} \label{16}%
\end{equation}%
\begin{equation}
(2\pi)^{4}|\lambda|^{2}m_{N}P(m_{N}^{2})=-B_{3}I_{1}(R)-B_{7}+a_{1}%
B_{9}+\Delta_{2} \label{17}%
\end{equation}

where%
\begin{equation}
I_{n}(R)=\int\limits_{0}^{R}dt\ t^{n}P(t) \label{17a}%
\end{equation}
and%
\begin{align*}
\Delta_{1}  &  =-(2\pi)^{4}\int\limits_{thr}^{R}dtP(t)\operatorname{Im}\Pi
_{1}(t)+a_{2}A_{10}\\
\Delta_{2}  &  =-(2\pi)^{4}\int\limits_{thr}^{R}dtP(t)\operatorname{Im}\Pi
_{2}(t)+a_{2}B_{11}%
\end{align*}
\ $A_{10}$ and $B_{11}$ are higher dimensional condensates. The integrals
appearing in the expressions of $\Delta_{1}$ and $\Delta_{2}$ as well as the
higher order condensates are unknown. Whereas it is possible to assess the
latter using the method of Pad\'{e} approximants the former are impossible to
estimate in practice. The only thing we can do is minimize these integrals and
then neglect them.

The choice of the function $P(t)$ aims at reducing this contribution as much
as possible in order to allow its neglect. This is achieved by minimizing
$P(t)$ over the resonance region. In the vast domain of QCD sum rules the
usual choice would be $P(t)=\exp(-t/M^{2})$ where the magnitude of $M$ (the
Borel mass) determines the strength of the damping of the contribution of the
continuum. If $M$ is small the damping is good but the contribution of the
unknown terms in the QCD asymptotic expansion of the amplitudes increases
rapidly. If $M$ increases the contribution of the unknown terms decreases but
the damping worsens. An intermediate value of $M$ has to be chosen from
stability conditions which are not met in the nucleon problem.

More information can be obtained if one uses $tP(t)$ as an integration kernel.
Here however one has to verify the validity of the neglect of the unknown
integrals which enter in $\Delta_{1,2}$ because the relative damping
$(tP(t)/m_{N}^{2}P(m_{N}^{2}))$ worsens. This quantity is still small in our
case whereas damping practically disappears in the case of the exponential kernel.

The value of \ $R$ should not be too small as this will invalidate the use of
the OPE on the circle, nor should it be too large as the kernel will start
enhancing the contribution of the continuum instead of suppressing it. An
intermediate value of $R$ around which the sum rules are stable should be chosen.

We then get%
\begin{equation}
(2\pi)^{4}|\lambda|^{2}m_{N}^{2}P(m_{N}^{2})=-A_{0}I_{3}(R)-A_{4}%
I_{1}(R)+A_{8}+\Delta_{1}^{\prime} \label{18}%
\end{equation}%
\begin{equation}
(2\pi)^{4}|\lambda|^{2}m_{N}^{3}P(m_{N}^{2})=-B_{3}I_{2}(R)-B_{9}+\Delta
_{2}^{\prime}, \label{19}%
\end{equation}
$_{{}}$where%
\begin{align*}
\Delta_{1}^{\prime}  &  =-(2\pi)^{4}\int\limits_{thr}^{R}%
dt\ tP(t)\operatorname{Im}\Pi_{1}(t)+a_{1}A_{10}+A_{12}\\
\Delta_{2}^{\prime}  &  =-(2\pi)^{4}\int\limits_{thr}^{R}%
dt\ tP(t)\operatorname{Im}\Pi_{2}(t)+a_{1}B_{11}+B_{13}%
\end{align*}

The nucleon mass can be determined by taking various ratios between
eqs.(\ref{16}) to (\ref{19}). Using $\left\langle \alpha GG\right\rangle
=.012\ GeV^{4}$, $\langle qq\rangle=-(1.\,90\pm0.14)\times10^{-2}GeV^{3}$ at
$\mu=2GeV$ \cite{CAD09} $\alpha_{s}(m_{\tau}^{2})=0.329\pm0130$\cite{Pich2013}
and neglecting the $\Delta^{\prime}s$ we obtain three expressions for the
nucleon mass%
\begin{equation}
m_{N}^{2}=(A_{0}I_{3}(R)+A_{4}I_{1}(R)+A_{8})/(A_{0}I_{2}(R)+A_{4}%
I_{0}(R)+A_{6}-a_{1}A_{8} \label{20}%
\end{equation}%
\begin{equation}
m_{N}^{2}=(B_{3}I_{2}(R)+B_{9})/(B_{3}I_{1}(R)+B_{7}-a_{1}B_{9}) \label{21}%
\end{equation}%
\begin{equation}
m_{N}=(B_{3}I_{1}(R)+B_{7}-a_{1}B_{9})/(A_{0}I_{2}(R)+A_{4}I_{0}%
(R)+A_{6}-a_{1}A_{8} \label{22}%
\end{equation}
At this point it is worth assessing the reliability of eqs.(\ref{16}) and
(\ref{17}) (and consequently of eqs.(\ref{18}) and (\ref{19}%
)\cite{Nasrallah2013}). It can be argued \cite{Leinweber} that eq.(\ref{17})
is more reliable than eq.(\ref{16}) because the first order radiative
correction to $A_{0}$ is anomalously large which casts doubt on the validity
of the QCD expansion. Another reason is that the dominant term is provided by
$A_{6}$ proportional to the 4-quark condensate the deviation of the value of
which from the one given by factorization being really unknown. We therefore
consider eq.(\ref{21}) to be the most reliable of the three equations above
and adopt it in our analysis. Of course factorization has still to be used to
evaluate $B_{7}$ and \ $B_{9}$ but these are now non -dominant terms.

Both numerator and denominator of eq.(\ref{21}) are stable in the interval
$2<t<3$ (see Fig.2 ) and so is the resulting value of the mass.%

\includegraphics{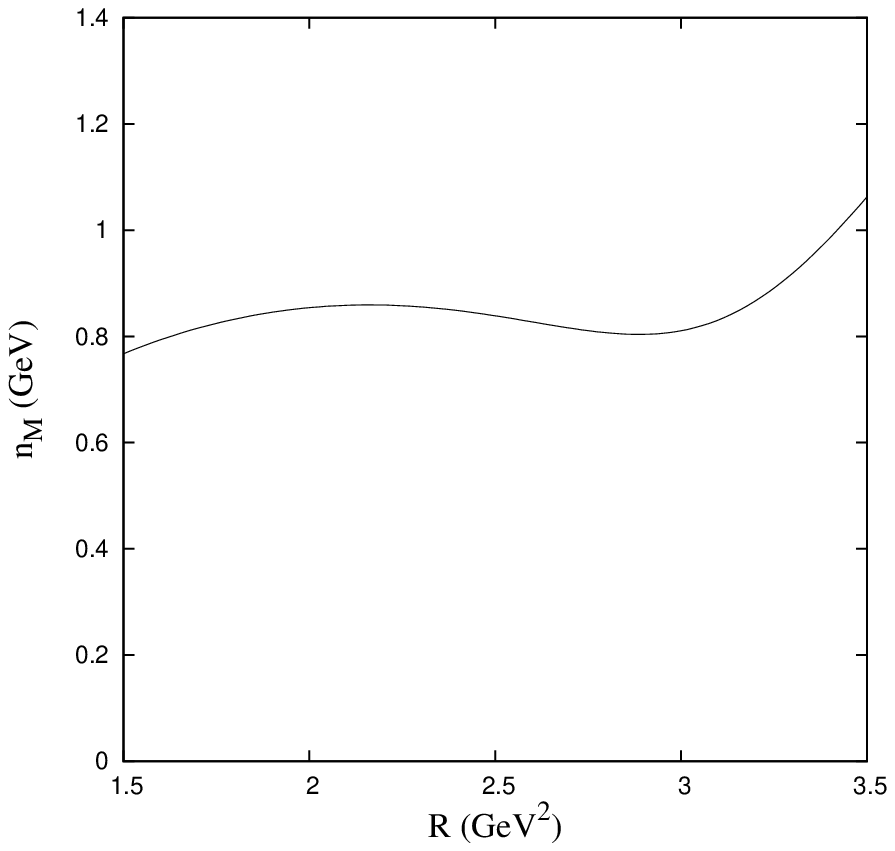}

Using standard values for the condensates eq.(\ref{21}) then yields\
\begin{equation}
m_{N}=(830\pm50)MeV \label{23}%
\end{equation}

The error having been obtained by estimating the higher order condensates
using Pad\'{e} approximants and by varying the coefficients $a_{1}$ and
$a_{2}$ within reasonable limits in order to deplete \ the contribution of the
continuum to the dispersion integrals.

If we use eqs.(\ref{20}) and (\ref{21}) nevertheless with the standard value
for $A_{0}$ and with $A_{6}$ obtained from factorization we get\
\begin{equation}
m_{N}=810MeV \label{24}%
\end{equation}
and
\begin{equation}
m_{N}=870MeV \label{25}%
\end{equation}
respectively.

The fact that all three ratios give almost the same answer pleads for the
choice of the standard values of \ the condensates.and for the smallness of
the contribution of instantons which we have neglected. An interesting study
of the effects of instantons is undertaken in \cite{Forkel}.The damping
ratio=value of kernel at resonance/value of kernel at nucleon shows that we
have it much better than exponential damping which in addition cannot take
moments (integral of $tP(t)$) and is not even stable.

\section{Results and conclusions}

With standard values of the condensates and using the factorization
approximation as discussed above, we get a value for the nucleon mass
$m_{N}=(830\pm50)MeV$ close to the experimental value. For a more precise
statement, we need do discuss possible errors in the sum rules. It is
important to distinguish two kinds of errors, theoretical and experimental. A
theoretical error of $\pm.03GeV$ arises from the uncertainty in the
condensates. We used $\left\langle \alpha GG\right\rangle =.012\ GeV^{4}$,
$\langle qq\rangle=-(1.\,90\pm0.14)\times10^{-2}GeV^{3}$ at $\mu=2GeV$
corresponding to the limits set by $m_{u}+m_{d}=(8.2\pm0.6)MeV$ \ (at scale
$\mu=2GeV$) \cite{CAD09} in the GMOR relation. Varying the scale parameter
$\mu^{2}$ between $4GeV^{2}$ and $2GeV^{2}$ introduces an additional error of
$\pm0.03GeV$. Furthermore there is an error due to the strong coupling
constant $\alpha_{s}(m_{\tau}^{2})=0.329\pm013$ \cite{Pich2013} which leads to
an error of $\pm0.6MeV$ . The total error in the calculated nucleon mass is
therefore mainly due to the method, i. e. due to the fact that the continuum
is not completely eliminated by our method.

Noting that our nucleon mass $m_{N}$ is obtained in the chiral limit, we get
for the total nucleon mass, which we denote by $M_{N}$%
\[
M_{N}=m_{N}+\sigma_{\pi N}+\sigma_{s}%
\]
To get some qualitative conclusions we assume that the strangeness content of
the nucleon can be estimated by the Zweig rule (or large $N_{C}$ )to be
$y=0.2$. If we use this result and additional theoretical prejudice,%

\[
\frac{m_{s}}{\hat{m}}\sim25,\text{ \ \ \ }\sigma_{\pi N}\sim45MeV\text{
\ Chiral Perturbation Theory\cite{Gasser}}%
\]
we can obtain an estimate for the strange sigma term
\[
\sigma_{s}=\frac{y}{2}\frac{m_{s}}{\hat{m}}\sigma_{\pi N}=2.5\sigma_{\pi
N}=113\ MeV
\]
which leads then to the final prediction of the nucleon mass
\[
M_{N}=990\pm50MeV
\]
with additional systematic errors arising from the uncertainties in the sigma
terms. Although errors are large, our result for the nucleon strongly prefers
a relatively small sigma term $\sigma_{\pi N}$ of order $45MeV$. There is also
strong indication that the strange content of the nucleon should be somewhat
smaller than $y=0.2$.

Alternatively we could use our result for $m_{N}$ together with eq.(\ref{2})
and a given nucleon sigma term to extract the strangeness content of the
nucleon. We find%
\[%
\begin{array}
[c]{c}%
\text{For }\sigma_{\pi N}=45MeV\text{\ \ \ \ \ }\Rightarrow\ y=0.11\\
\text{For }\sigma_{\pi N}=80MeV\text{\ \ \ \ \ }\Rightarrow\ y=.03
\end{array}
\]

In conclusion, we have presented a sum rule calculation of the nucleon mass in
the chiral limit using a kernel in the dispersion integral tailored to
minimize the contribution of the unknown continuum without involving the
higher order unknown condensates. Using standard values of the known
condensates and the sigma terms, we obtain for $M_{N}$ a value which agrees
quite well with the experimental one and \ which excludes a large value of the
strageness content of the nucleon.

Acknowledgement:

This work was started when the authors were visiting the American University
of Beirut, Lebanon. We like to thank the Physics Department for its hospitality.

\end{document}